# Fast response deep-ultraviolet photodetector based on high-quality single-crystalline CVD diamond

Le Li

University of Chinese Academy of Sciences, 19 Yuquan Road, Beijing 100049, China

A high-performance fast response deep ultraviolet (UV) photodetectors with interdigitated Ti/Au planar electrodes using a lithography technology has fabricated on homoepitaxial diamond. The device shows a high ultraviolet photocurrent at 213 nm, which is eight orders of magnitude higher than the dark current at the bias voltages of 30 V. In addition, the time-resolved photoresponse measurements using a mechanical method and a pulsed 213 nm laser show that a good cycling, low persistent photoconductivity, and transient time-resolved response of up to 2.4 ns.

1 Introduction

In the past decades, diamond has attracted much attention as the host material of solar-blind ultraviolet (UV) photodetectors (PDs), owing to its wide band-gap (5.5 eV), fascinating mechanical property, ultrahigh carrier mobility (4500 cm$^2$ / (V · s) for electrons and 3800 cm$^2$ / (V ·s) for holes)[1-2], remarkable thermal stability and highest thermal conductivity (22Wcm$^{-1}$K$^{-1}$).[3]Especially with rapidly development of the synthesis technology, chemical vapor deposition (CVD) epitaxial single crystal diamond has widely used in deep UV photodetector to environmental security, information technology, medical treatment, and astronomical observation, which are of great importance in both civilian and military fields[4-5]. For an ideal deep UV photoconductor, it should display high light response capability in the UV region, high signal-to-noise ratio, good visible rejection (i.e, high spectral selectivity) and suitable for high radiation as well as harsh chemical environment conditions.

Nowadays, the researchers have been making efforts in improving the performance of diamond deep UV detectors by designing various device structures and different electrode geometries[6-9], Alvarez et al. reported the metal-semiconductor-metal (MSM) Shottky photodiode fabricated on as-grown homoexpitaxial diamond thin film with hydrogen-terminated surface.[10] Nesladek reported a pin-structure diamond photodiode on a large-area n-type polycrystalline CVD diamond used for detectors or electron emitters. [11] Liu et al. applied three dimensional metal electrodes in diamond UV detector, and that a strong persistent photoconductivity (PPC) appeared. [12] Only recently K. Liu et al. fabricated a detector with a groove-shape electrode, while the responsivity was rather low.[13] Currently the diamond photodetectors still have difficulty in meeting the perfect 5s [14] requirements due to the more defect density in diamond and non-diamond impurities.

In this article, we aim to fabricate high-performance diamond photodetectors with high sensitivity, excellent photo responsivity, fast response time, and low noise on high quality homoepitaxial un-doped diamond film with oxidized surface.

2. Experimental details

The single crystal diamond was homoepitaxially grown on (100)-oriented $5\times5\times0.42mm^3$ high-pressure high-temperature Ib synthetic (100) diamond substrate using a microwave plasma-enhanced CVD method. The concentration ratio of $CH_4$, the temperature, the growth pressure, microwave power were 1200 ℃, 170 Torr, 8% and 5000 W , respectively. After growth, the sample was boiled in acid mixture ($H_2SO_4$:$HNO_3$ = 1:1 by volume) at 350 °C for 2h, then oxygen plasma treatment was used for 5minites to change the hydrogen terminated surface to oxygenated surface. About 5 nm thick titanium (Ti) electrodes followed by a 50nm Au were covered on the diamond surface using standard lithography technology and DC magnetron sputtering method to give a planar device comprising interdigitated electrodes. The width of electrode was 10μm, which is the same with the distance between two electrodes, leading to a total active area of about 0.0528 $mm^2$.

The optoelectronic characterization was evaluated in air with a Keithley2450 voltage source. The deep UV incident wavelength of 213nm was provided by a deep ultraviolet light spectrometer. Time response behaviors were measured by putting the light on and off through a metal shutter. The transient signal was recorded by a 213nm pulsed laser with repetition frequency of 10-60Hz device , which is also known as the time of flight (ToF) technique in which the duration of the current pulse read on a oscilloscope by the drift of free charge carriers under the influence of an externally applied electric field.

3. Results and discussion

In deep UV detection devices, the performance largely depends on the quality of the diamond. Cathodoluminescence (CL) spectrometer is an important characterization method to clarify the crystalline quality from the viewpoint of electronic states. The crystalline quality of the sample as well as the impurity defects are investigated by exciting the luminescent center of single crystal diamond. Figure 1 shows the CL spectrum of the homoepitaxial diamond layer. Clear edge emissions can be observed form 235nm to 245nm, and the fine structure. The strongest peaks at 235 nm is due to intrinsic emissions attributed to free-exciton recombination radiation associated with a transverse optical phonon. CVD single crystal diamond has a high crystalline quality.

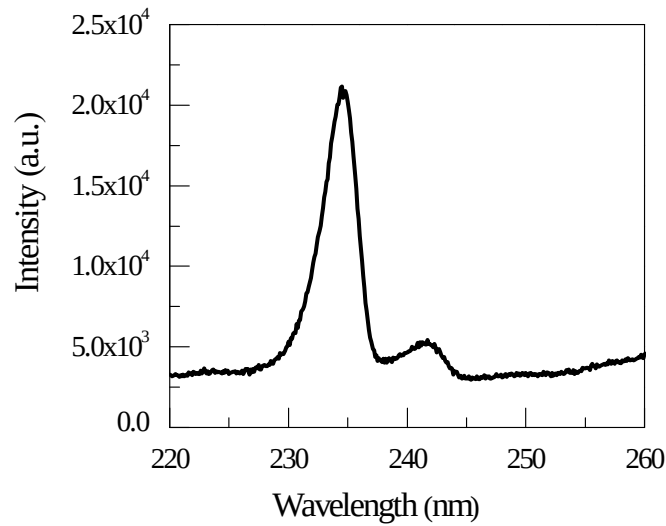

Figure 1. RT CL spectra of a single crystal CVD diamond

Here the intrinsic diamond we grown means homoepitaxial diamond layer unintentional doping. Due to the high resistivity of the epilayers, the devices exhibited low dark current. The Fig. 2 shows the dark current-voltage (I-V) characteristics of a photodetector fabricated on the intrinsic diamond epilayer. The dark current is about $3\times10^{-12}$ A at 30 V, The low dark current is consistent with previous works [15-16] where the photoelectric detection devices with a very low leakage current at $10^{-12}$A. Figure 3 (a) (b) (c) show the I-V characteristics under the illumination of 213 nm, whose power densities are 20 mW/mm$^2$, 200 μW/mm$^2$ and 5 μW/mm$^2$, respectively. When the 20 mW/mm$^2$, 200 μW/mm$^2$ and 5 μW/mm$^2$ light densities used, the current at 30 V are $1.6\times10^{-4}$A, $1.5\times10^{-6}$A, $2\times10^{-7}$A, respectively. When the device was under the light power intensity is 20mW/mm$^2$, the photocurrent rapidly increases from $6.5\times10^{-6}$ A to $1.9\times10^{-5}$ A with a small voltage of 0.1 V. The photocurrent increases eight orders of magnitude higher than the dark current and reaches a saturation value close to $10^{-4}$A for bias at 3.8V. The increase of photocurrent is consistent with the above phenomenon. However, with the decrease of light intensity, the need of photocurrent saturation voltage increased.

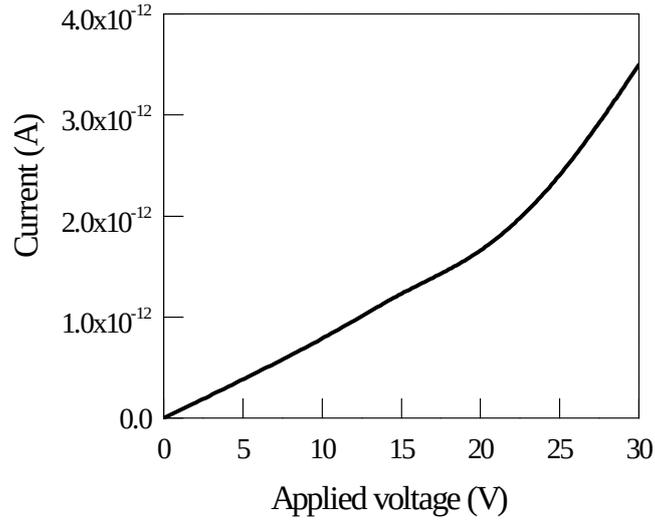

Figure2. Dark I-V characteristics of the device

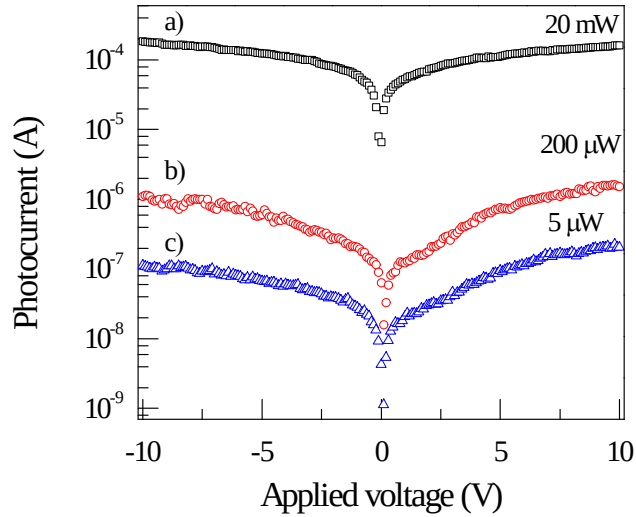

Figure3. Photocurrent at 213 nm vs applied voltage measured at light intensity (a) 20 mW/mm²; (b) 200 μW/mm² and (c) 5 μW/mm²

For a photoconductor, the responsivity (R) is a performance metric for characterizing photodetectors, which is defined as the ratio of the photocurrent due to optical signals to the incident optical power. R can be expressed by the following equation:

$$R = \frac{I_{ph}}{P_{opt}} \tag{1}$$

Where $I_{ph}$ is the photocurrent, and $P_{opt}$ is the incident optical power.

Figure 4 (a) shows the responsivity of the sample varying with the bias voltage increasing under 5μW/mm² light power intensity. As a result, with the applied voltage increased, the responsivity increases rapidly and the saturates reaching a large value of 13A/W. We found that photodetectors can detect weak light sensitively, as the intensity of incident light became weaker, the responsiveness is higher. The responsivity of the sample detector is calculated to be 1.4A/W and 2.67 A/W with 20mW/mm² and 200µW/mm², respectively.

Another important parameter for measuring diamond detectors is quantum efficiency(QE), which is the number of electron-hole pairs converted by the incident photons with frequency $\bullet$. The external quantum efficiency can be describe as $QE = \frac{I_{ph}}{q\psi}$ ,where ψ is a photon flux that equals ${P_{opt}}/{h\upsilon}$ ,so the quantum efficiency curves are similar to the responsivity plots. Figure 4 (b) shows the external quantum efficiency of the device varying with the bias voltage increasing under 5µW/mm² light power intensity. When electric field E of 3×10⁴ V/cm is applied across the device incident illumination intensity at 213 nm are 20mW/mm², 200µW/mm² and 5µW/mm², corresponding to the quantum efficiency of 828%, 760% and 7560%, respectively.

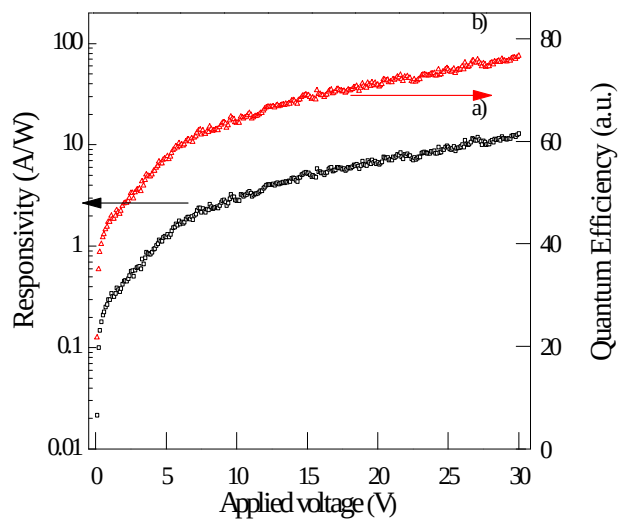

Figure.4 (a) Current responsivity (b) External quantum efficiency varying with the bias voltage increasing under 5µW/mm²

The detector time response behaviors on the applied voltage of 10V was illustrated in Figure 5, which was measured by putting the 213nm light on and off through a metal shutter, the switch interval is 10 seconds. When the light is on, photocurrent increases drastically, then increases a little slowly until the current up to 10⁻⁸A. When the light is off, the current drops quickly and tends to be stable at the

10⁻¹¹ A. During the test period, the device current circulating is strong, there is no attenuation, and the current can reach the original value, exhibiting there is no obvious persistent photoconductive conductivity, which may be contributed to the fewer defects in the diamond we prepared. Because the defects will trap carriers, resulting in the decay of rise time and fall time.[17]

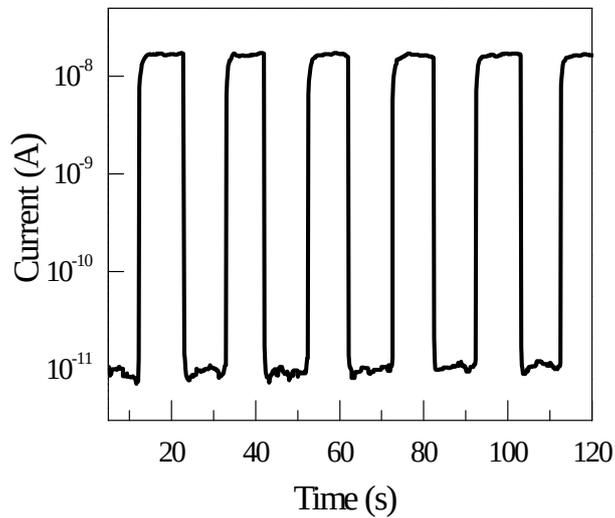

Figure. 5 Time response of the device under 213nm illumination. The light power density is 5μW/mm², and the applied bias is 10 V.

The fast transient response of single diamond deep-UV PDs in our work can be comprehended by analyzing the carrier transport processes. Figure 6 shows the transient response of the device which applied different bias voltage under the illumination pulse of 213 nm light. The light pulse is generated by a excimer laser, with an available frequency of 10-60 Hz and the pulse width of 1 ns. Besides, the signal becomes stronger as the voltage increases, the rise time and fall time are shortened, and then the full width at half maximum (FHWM) of the waveform becomes smaller. While dependence of the FWHM on the bias was not obvious, the response time of pulse current on a oscilloscope is about 2.4ns, as the voltage increases vary by 0.01 on the order of magnitude. Compared with the previously prepared photodetectors, this one shows much faster response times, and there is little loss of delay.

In the first approximation, by knowing the drift time t and distance d , the carrier drift velocity υ can be calculated as $υ = d/t$, at low field current does not reach saturation, $υ = μ_0 E$ . By changing the electric field, the time through the device changes, ₄t can be used to estimate the carrier mobility. The carrier mobility in the device was estimated approximately equal to about 3600 cm² / (V · s).  Accurate calculation of device carrier mobility, which is supposed by the quick response time as will be discussed later.

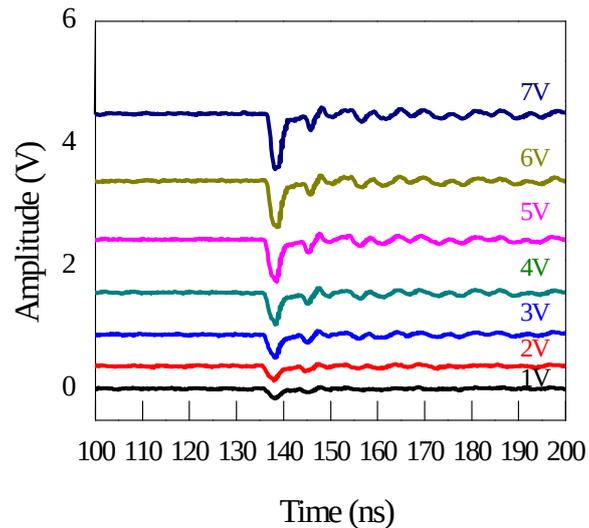

Figure 6 Transient time-resolved response of the photodetector at different voltages.

In summary, a fast response, high performance photodetector with the photocurrent discrimination ratio up to $10^8$ between illumination at 213nm and darkness was fabricated on an intrinsic homoepitaxial diamond. Due to its fewer defects, it has a high hole electron mobility. Therefore the photovoltaic detector has a high-speed operation. And the response time is at least 2.4 ns, the PPC was weaker than previously prepared photodetectors. We gain a high EQ of 75.6 and a responsivity of 13 A/W at 213nm light for a bias of 30 V. A further improvement in the mobility measurements of high-quality diamond.